\documentclass[sigconf]{acmart}

\usepackage[english]{babel}
\usepackage{tikz}
\usepackage{amsmath}
\usepackage{filecontents}
\usepackage{hyperref}
\usepackage{enumitem}
\usepackage{graphicx}
\usepackage{subcaption}
\usepackage{url}
\usepackage{algorithmicx}
\usepackage{algpseudocode}
\usepackage{algorithm}
\usepackage{multirow}
\usepackage{makecell}
\usepackage{appendix}

\AtBeginDocument{
  \setlength{\abovedisplayskip}{3pt plus 1pt minus 1pt}
  \setlength{\belowdisplayskip}{3pt plus 1pt minus 1pt}
  \setlength{\abovedisplayshortskip}{0pt plus 1pt}
  \setlength{\belowdisplayshortskip}{2pt plus 1pt}
}

\setlist{nosep, leftmargin=*}

\renewcommand\footnotetextcopyrightpermission[1]{} 
\setcopyright{none}

\acmDOI{}

\acmISBN{}

\acmPrice{}

\begin{document}

\title{SAIL: Perceptual Quality-Aware Rate Control for Cloud Gaming}

\author{Houde Qian}
\authornote{Both authors contributed equally to this research.}
\affiliation{%
  \institution{Tsinghua University}
}
\email{qhd23@mails.tsinghua.edu.cn}

\author{Chenglei Wu}
\authornotemark[1]
\affiliation{%
  \institution{Tencent}
}
\email{wuchenglei1993@gmail.com}

\author{Jiaxing Zhang}
\affiliation{%
  \institution{Tencent}
}
\email{zhangjiaxing1998@gmail.com}

\author{Rui-Xiao Zhang}
\affiliation{%
  \institution{University of Illinois Urbana-Champaign}
}
\email{zrx389139970@gmail.com}

\author{Jing Wang}
\affiliation{%
  \institution{Tencent}
}
\email{jingwang0814@gmail.com}

\author{Meijia Song}
\affiliation{%
  \institution{Tencent}
}
\email{s122501490@gmail.com}

\author{Sijia Chen}
\affiliation{%
  \institution{Tencent}
}
\email{sijiachen@tencent.com}

\author{Xiaozhong Xu}
\affiliation{%
  \institution{Tencent}
}
\email{xiaozhongxu@tencent.com}

\author{Zhi Wang}
\affiliation{%
  \institution{Shenzhen International Graduate School, Tsinghua University}
}
\email{wangzhi@sz.tsinghua.edu.cn}

\author{Lifeng Sun}
\authornote{Corresponding authors.}
\affiliation{%
  \institution{Tsinghua University}
}
\email{sunlf@tsinghua.edu.cn}

\author{Honghao Liu}
\authornotemark[2]
\affiliation{%
  \institution{Tencent}
}
\email{hritian@gmail.com}

\renewcommand{\shortauthors}{Qian et al.}

\begin{abstract}
Cloud gaming streams cloud-rendered frames under strict motion-to-photon latency, yet its at-scale viability is increasingly constrained by bandwidth cost: in our study of the $T$ cloud gaming platform, bandwidth accounts for 30--60\% of total operating expense. This high bandwidth consumption stems from a fidelity-first objective of making the stream perceptually indistinguishable from local gameplay. It drives production systems toward best-effort bitrate allocation that pushes the encoder to the highest rate allowed by congestion control. However, the bitrate-perception relationship saturates: beyond a frame-dependent \emph{perceptually lossless} threshold, additional bits yield negligible perceptual improvement, creating systematic \emph{redundant quality} that wastes bandwidth.

We present \textbf{SAIL}, a production quality-aware rate control system with the goal of achieving perceptually lossless quality while avoiding unnecessary bandwidth waste. SAIL adopts a \emph{post-encoding} architecture to enable millisecond-scale feedback at near-zero overhead. It comprises three key designs: (i) an encoder-driven quality assessment model that leverages zero-cost encoder outputs for real-time quality estimation; (ii) a hybrid rate control mechanism that balances steady-state adaptation with dynamic spike absorption; and (iii) a network-aware strategy that coordinates with congestion control to prevent capacity underestimation.
SAIL has been fully deployed on the $T$ cloud gaming platform and reduces bandwidth consumption by 44.27\% and end-to-end latency by 8.37\% without degrading perceived quality, serving tens of millions of users and accumulating billions of hours of total gameplay.
\end{abstract}

\maketitle

\section{Introduction}

Cloud gaming moves compute-intensive graphics rendering from resource-constrained end devices to cloud servers, enabling high-fidelity gameplay on low-end PCs and mobile phones. This architectural shift has attracted substantial industrial interest and investment, with market research projecting a global market size of USD 18.71 billion by 2027~\cite{market-survey}. Despite this growth, large-scale deployment remains cost-constrained, with bandwidth spending as one of the main contributors. In our analysis of the $T$ cloud gaming platform, bandwidth accounts for 30--60\% of total operating cost, directly limiting scalability and profitability.

This high bandwidth cost of cloud gaming is largely dictated by its fidelity objective: the cloud-rendered stream is expected to be visually indistinguishable from local gameplay. This objective, drives production systems toward best-effort bitrate allocation (\S~\ref{subsec:best-effort}), where the encoder is pushed to the highest rate permitted by congestion control.
For instance, NVIDIA GeForce NOW provisions 35--45~Mbps for 4K cloud gaming~\cite{nvidia-bitrate}.
However, higher encoding bitrates can not always yield better visual quality. Once beyond a certain threshold—\textit{perceptually lossless (p-lossless)} threshold, increasing encoding bitrate yields negligible improvements in visual quality, causing a substantial fraction of frames to carry redundant quality while still consuming additional bandwidth.
As evidenced in Figure~\ref{fig:system-overview}, scenes with minimal temporal dynamics may require a bitrate as low as $\sim$10~Mbps to maintain p-lossless quality.

These observations necessitate the adoption of Quality-Aware Rate Control (QARC), which dynamically caps the per-frame encoding bitrate at the \textit{p-lossless} threshold, thereby eliminating redundant bandwidth consumption while guaranteeing perceptual fidelity.
Realizing this vision in production, however, is impeded by four fundamental constraints intrinsic to cloud gaming architecture: \textbf{\emph{(i) an ultra-low latency budget}}, where pixel-domain Visual Quality Assessment (VQA) pipelines incur prohibitive inference delay\footnote{The processing latency of neural-network-based feature extractors often exceeds 100 ms.} that violates strict latency requirements; \textbf{\emph{(ii) the prohibitively high cost of per-session computation}}, as the one-to-one service model renders the use of high-end GPUs (e.g., NVIDIA GeForce RTX 4090) with limited throughput economically and operationally impractical; \textbf{\emph{(iii) tight coupling with congestion control}}, which dictates that rate changes must align with network availability; and \textbf{\emph{(iv) reliance on black-box hardware encoders}} that expose minimal control interfaces.

%

These constraints force us to pursue a QARC design that is simultaneously low-latency, low-overhead, and deployable in production cloud gaming.
However, realizing QARC in cloud gaming hinges on addressing three coupled challenges. \textit{\textbf{Challenge 1: low-latency, low-overhead VQA}}: the strict latency constraints of cloud gaming preclude any \textit{\textbf{pre-encoding}} analysis (e.g., pixel-domain VQA or lookahead planning) that resides on the motion-to-photon critical path and introduces unacceptable delays, while limited hardware interfaces restrict \textit{\textbf{in-loop}} quality analysis. Consequently, the only feasible approach is to provide accurate and reliable quality assessment through \textbf{\textit{post-encoding}} analysis derived from encoder outputs within a tight millisecond-scale latency budget; \textit{\textbf{Challenge 2: robust reactive rate control}}: the system must manage the inherent \textit{\textbf{one-frame control lag}} of the post-encoding analysis scheme, translating quality feedback into stable yet responsive next-frame bitrate decisions despite highly non-stationary bitrate-quality dynamics; and \textit{\textbf{Challenge 3: congestion control compatibility}}: coordinating with congestion control to prevent bandwidth underestimation during sustained under-sending, while ensuring rapid ramp-up during complexity spikes.



To address this critical gap, we propose SAIL, a QARC solution tailored for cloud gaming at scale. Specifically designed to meet the demands of large-scale deployment, SAIL achieves precise, real-time bitrate control through three synergistic components:

\noindent\textbf{(i) Encoder-Driven VQA (\S ~\ref{sec:design-vqa}).}
Through an in-depth analysis of encoder outputs, we identify specific statistics that exhibit a strong correlation with visual quality. Leveraging these intrinsic signals, we implement a lightweight VQA model that operates with negligible overhead. Furthermore, we adopt diverse optimization strategies to enhance the model's accuracy, thereby ensuring its reliability for real-time rate control.

\noindent\textbf{(ii) Hybrid Rate Control (\S ~\ref{sec:rate-control}).}
We establish our design principles through rigorous user studies to clarify and quantify the control target. Furthermore, through extensive measurements of cloud gaming sessions, we decompose the complex control problem into tractable components, enabling the design of a hybrid control framework that achieves substantial bandwidth savings without compromising visual quality.

\noindent\textbf{(iii) Network-Aware Co-Design (\S ~\ref{sec:system-level-optimization}).}
To reconcile bitrate savings with transport availability, SAIL actively coordinates with the congestion control layer. It employs a compensatory probing mechanism that combines Bandwidth Utilization Ratio (BUR)~\footnote{BUR is derived from Pudica~\cite{Wang2024}, a state-of-the-art congestion control algorithm for cloud gaming.} and network capacity assessments to prevent the transmission window from shrinking during low-bitrate periods, ensuring immediate responsiveness when quality demands recover.

\begin{figure}[t!]
    \centering
    \includegraphics[width=\linewidth]{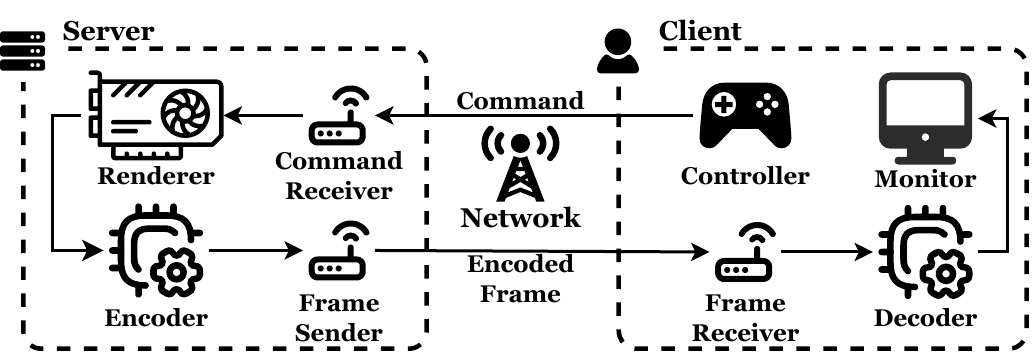}
    \caption{A typical cloud gaming system.}
    \label{fig:cg}
\end{figure}

Our contributions are as follows:

\begin{itemize}[wide, topsep=0pt, itemsep=0pt]

\item By analyzing a large-scale commercial cloud gaming system, we are the first to quantify the substantial cost of redundant quality inherent in conventional best-effort bitrate allocation.


\item We present SAIL, a practical QARC system for cloud gaming that realizes a post-encoding, frame-level control loop via (i) an encoder-driven lightweight VQA, (ii) a hybrid rate controller, and (iii) a network-aware co-design.

\item We implemented SAIL using the NVENC~\cite{nvenc} and demonstrated its effectiveness through comprehensive evaluations. In real-world deployment, SAIL reduces bandwidth consumption by 44.27\%, decreases the latency by 8.37\%, and has supported over 20 million cloud gaming sessions, proving its practical viability.
\end{itemize}

\noindent\textbf{Ethical claim.} 
All user data collected in this work are obtained with explicit permission from the users and properly anonymized to protect their privacy. This work does not raise any ethical concerns and conforms to the IRB policies of the authors' institutions.

\section{Background and Motivation} \label{sec:background-and-motivations}

\subsection{Best-Effort
Bitrate Allocation in Cloud Gaming System}
\label{subsec:best-effort}

In modern cloud gaming systems, user inputs are sent to a remote server, where the game's visual output is encoded into a video stream and delivered to the client for real-time decoding and display (Figure~\ref{fig:cg}). Unlike conventional real-time video applications (e.g., video conferencing), players inevitably compare cloud gaming with the local gameplay experience, imposing stringent requirements on both \emph{visual quality} and \emph{E2E latency}. On one hand, real-time visual quality must be near-indistinguishable from local rendering; on the other hand, user engagement exhibits high sensitivity to latency, with even a 10~ms increase leading to substantial engagement loss, which requires the E2E latency to be strictly controlled within 20~ms~\cite{Wang2024}.

Despite substantial progress in latency reduction with continuous congestion control~\cite{Wang2024,ray2022sqp,jia2024tackling} and buffering optimizations~\cite{zhao2024jitbright,huang2025ace}, existing systems provide limited explicit control over visual quality.
Instead, to meet stringent user requirements for visual quality, these systems typically adopt a best-effort bitrate allocation strategy, driving the encoding bitrate to the maximum level permitted by congestion control in an effort to optimize visual fidelity.
In practice, to prevent unbounded rate inflation, the server configures a conservative bitrate cap $R_c$ as an upper bound on the encoder's output rate. Consequently, the realized encoding bitrate $R$ is jointly determined by the cap and the estimated available bandwidth $C$, given by $R = \min(R_c, C)$.

\begin{figure}
\centering
\includegraphics[width=\linewidth]{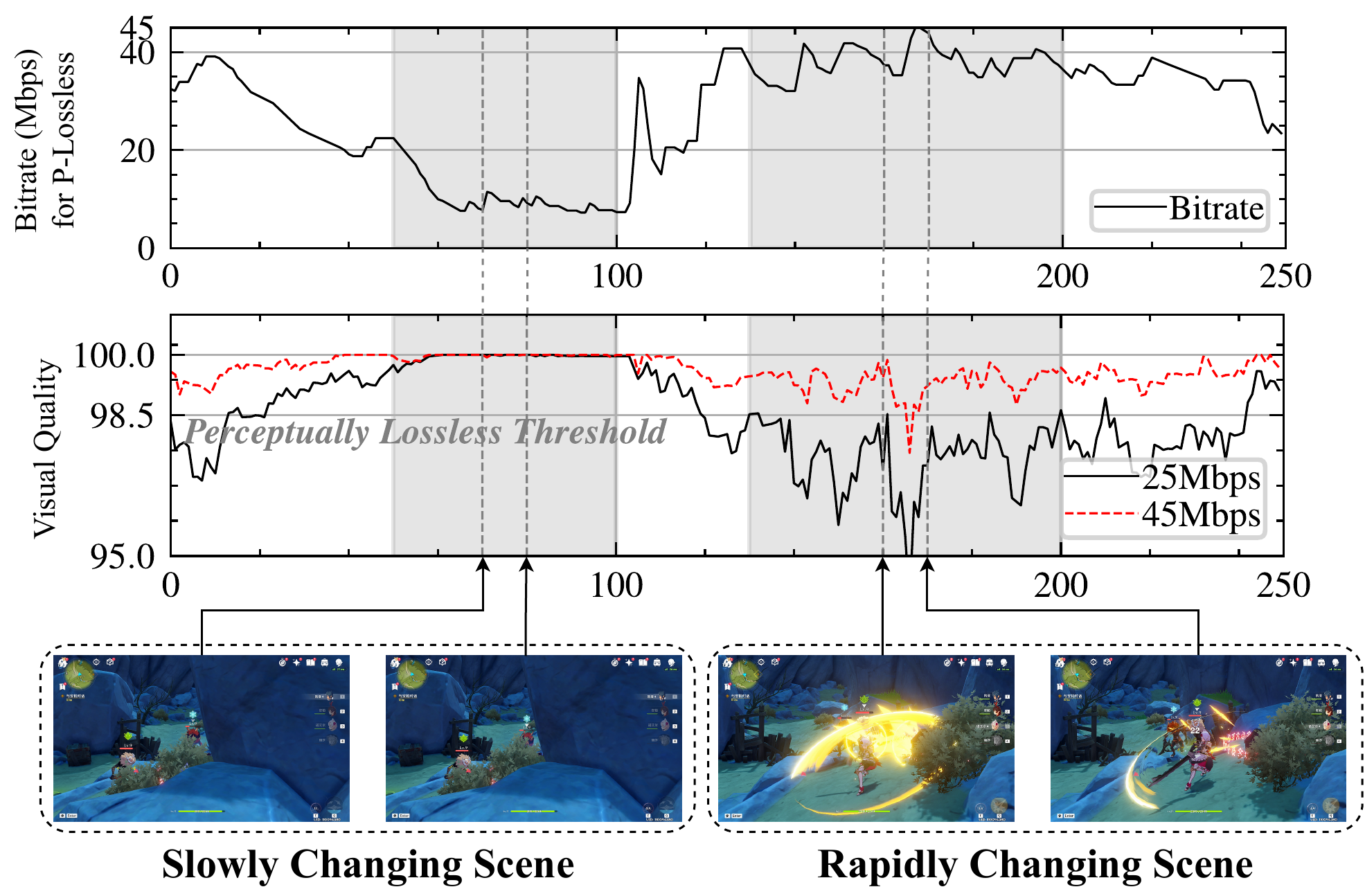}
\caption{Comparison of video quality under different encoding bitrates. The bitrate required to achieve p-lossless quality varies significantly across scenes, demonstrating that QARC can achieve substantial resource savings by eliminating perceptually negligible bitrate allocation.}
\label{fig:system-overview}
\end{figure}

This design, while widely adopted, suffers from \textbf{\textit{bitrate-quality mismatch}}: A constant $R_c$ fails to accommodate the extreme variance in cloud gaming content. For simple scenes, the p-lossless threshold is often far below $R_c$, leading to substantial bandwidth waste. Conversely, complex scenes may require bitrates far exceeding $R_c$ to maintain quality, yet the static cap prevents sufficient bit allocation, causing perceptible quality drops. As shown in Figure~\ref{fig:system-overview}, while 99.5\% of frames have already reached quality saturation at 45~Mbps and simple scenes may saturate at bitrates as low as 10~Mbps, rapidly changing scenes often require significantly more than 45~Mbps to achieve perceptual losslessness, yet are constrained by the static cap.
Moreover, higher encoded rates enlarge frame sizes and prolong transmission, increasing per-frame delivery delay and inflating operational costs, especially under bandwidth fluctuations.


\subsection{Opportunities and Incompatibilities: Quality-Aware Rate Control}

These limitations mentioned above motivate QARC as a promising alternative to best-effort bitrate allocation. Instead of imposing a single conservative cap $R_c$ across all frames, QARC sets a content-adaptive cap $R_{c,i}$ for each frame $i$ (e.g., via visual complexity), with the goal of achieving p-lossless quality while avoiding unnecessary bandwidth waste.



Although QARC has been widely adopted in Video-on-Demand (VoD) and live streaming, these domains operate under fundamentally different assumptions. In such systems, encoding is decoupled from playback, meaning that encoding latency does not gate the user's viewing experience. Moreover, content is typically encoded once and consumed by millions (the \textit{one-to-many} model), allowing high computational overhead to be amortized. Furthermore, the encoding bitrate is independent of real-time network throughput, isolating the encoder from congestion control. Consequently, traditional QARC solutions predominantly employ \textit{batch-based} and \textit{offline} (or at least \textit{semi-offline}) workflows~\cite{huang2019comyco, qin2019quality}: they buffer batches of frames for intensive complexity analysis before allocating bitrates. While effective for VoD and live streaming, these methods incur latency and computational costs that are prohibitive for highly interactive applications.\footnote{While some Real-Time Communication (RTC) solutions~\cite{Huang2018, 10.1145/3474085.3475594} attempt to avoid lookahead by predicting bitrates from historical observations, they assume a stable bitrate-quality relationship--a premise violated by the motion-conditioned, frame-level complexity jumps inherent in interactive cloud gaming.}

In contrast, cloud gaming embeds video encoding deeply within the real-time video generation pipeline, imposing four critical constraints that render traditional approaches infeasible:

\noindent\textbf{(i) Ultra-Low Latency Budget.} Cloud gaming requires extremely low encoding latency, as it sits directly on the critical path of E2E system latency. Any lookahead or heavyweight analysis immediately degrades interactivity.

\noindent\textbf{(ii) Economic Sensitivity to Computation.} As a \textit{one-to-one} real-time service where each frame is encoded and consumed only once, cloud gaming cannot amortize encoding costs. Any additional per-frame computation linearly increases the operational cost of large-scale deployment.

\noindent\textbf{(iii) Coupling with Congestion Control.} To guarantee real-time performance, frames must be transmitted immediately after rendering. This tight coupling compels rate control to function as the actuation mechanism for congestion control, dynamically adjusting output to match network capacity.

\noindent\textbf{(iv) Hardware Constraints.} Large-scale platforms rely on dedicated hardware encoders (e.g., NVENC) co-located on the rendering GPU to minimize latency and cost. However, these encoders expose limited control interfaces (e.g., target bitrate), precluding the fine-grained optimization typical of software encoders.

These strict constraints, unique to cloud gaming, are difficult to satisfy with traditional schemes, motivating the urgent need for a lightweight, frame-level, system-level QARC solution with near-zero additional latency.

\begin{figure}[t!]
    \centering
    \includegraphics[width=\linewidth]{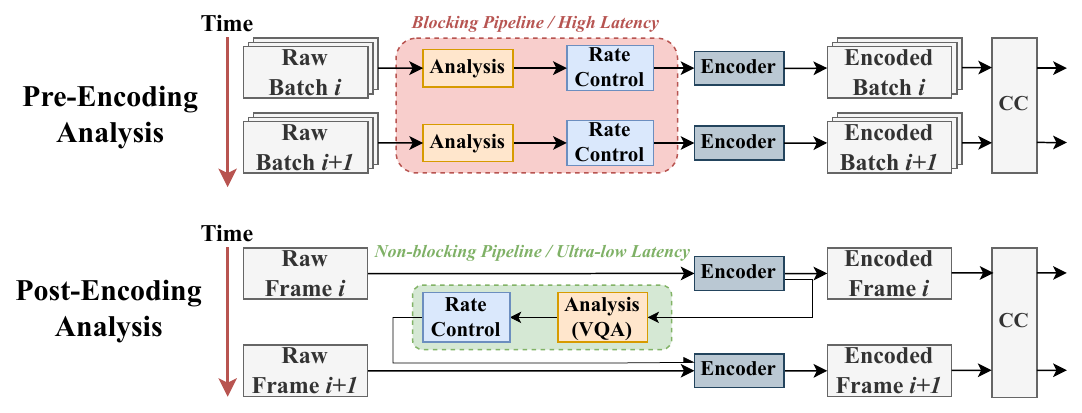}
    \caption{QARC architecture under different scenarios.}
    \label{fig:qas-architectures}
\end{figure}

\section{Cloud Gaming Tailored QARC}

Given QARC's promise for \textbf{\emph{bandwidth conservation}}, \textbf{\emph{quality improvement}}, and \textbf{\emph{latency reduction}}, we are motivated to explore a deployable QARC solution for cloud gaming at a large scale.
As illustrated in Figure~\ref{fig:qas-architectures}, the QARC workflow comprises three essential components: (i) visual quality analysis module, (ii) rate control module based on analysis results, and (iii) Congestion Control (CC) tailored to the decoupling of encoding bitrate and network capacity. However, deploying this workflow in production systems is non-trivial, as the stringent requirements and hardware constraints of cloud gaming impose significant challenges.

\subsection{Challenges of Realizing QARC in Cloud Gaming}

\subsubsection{Challenge 1: Low-Latency and Low-Overhead VQA}
Achieving accurate VQA typically demands computationally intensive pixel-level analysis, ranging from industry standards such as VMAF~\cite{vmaf} to advanced neural network-based approaches~\cite{kim2018deep, liu2024tempcompass}.
These methods incur substantial overhead and latency, failing to meet the stringent real-time requirements of cloud gaming.
As illustrated in Figure~\ref{fig:vqa-latency}, we benchmarked the full VMAF v0.6.1 implementation and a basic Convolutional Neural Network (CNN)-based feature extractor on an NVIDIA GeForce RTX 4070 GPU, measuring processing times over $10^4$ 1080p frames. The results demonstrate that both methods require at least 15~ms per frame, failing to satisfy the strict sub-10~ms latency budget essential for cloud gaming. Furthermore, their latency distributions are long-tailed, indicating the potential for processing bursts that could compromise system stability.
Given that feature extraction alone exhausts the E2E latency allowance, placing quality assessment on the critical path prior to encoding is infeasible. Instead, it necessitates a \textbf{\textit{post-encoding}} approach based on \textit{\textbf{encoder outputs}}, as illustrated in Figure~\ref{fig:qas-architectures}. However, large-scale cloud gaming systems rely on specific hardware encoders that expose limited control interfaces and narrow operational flexibility. Consequently, achieving low-latency, low-overhead, and precise VQA solely from hardware encoder outputs represents a significant and unaddressed challenge.

\subsubsection{Challenge 2: Robust Reactive Rate Control}
Hardware encoders expose video statistics only upon the completion of frame encoding, necessitating retrospective quality assessment. This limitation introduces an inherent \textit{\textbf{one-frame control lag}}, as the bitrate decision for the current frame relies entirely on feedback from the previous one. This latency poses a significant challenge for robust control. During abrupt spikes in visual complexity (e.g., rapid user-driven motion), initial frames are inevitably encoded using outdated bitrate parameters, leading to transient quality degradation. This issue is exacerbated by the highly non-linear and non-stationary nature of the bitrate-quality relationship in cloud gaming, where static heuristics or history-based regressors frequently result in under-reaction, overshoot, or oscillation. Consequently, a deployable post-encoding controller must ensure prompt detection of and swift recovery from quality drops, employing a robust mapping strategy that balances rapid correction with control stability.


\subsubsection{Challenge 3: Congestion Control Compatibility} Congestion control in cloud gaming is typically implemented by directly regulating the video encoding bitrate~\cite{Wang2024}. But QARC fundamentally alters the sending dynamics observed by congestion control. During simple scenes, bitrate reduction can induce sustained under-sending, prompting CC to downshift its bandwidth estimate and lose track of available headroom. When a subsequent complexity spike demands a rapid rate increase, CC may throttle the encoding bitrate based on the degraded estimate. Mitigating this is essential to preserve both low latency and accurate bandwidth estimation.

\begin{figure}[t]
    \begin{minipage}[t]{0.48\linewidth}
        \includegraphics[width=\linewidth]{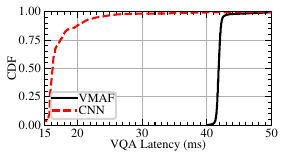}
        \caption{Computational latency of VMAF/CNN.}
        \label{fig:vqa-latency}
    \end{minipage}
    \hfill
    \begin{minipage}[t]{0.48\linewidth}
        \includegraphics[width=\linewidth]{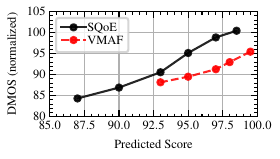}
        \caption{User ratings vs. predicted SQoE/VMAF scores for gaming content.}
        \label{fig:vmaf-and-sqoe}
    \end{minipage}
\end{figure}

\subsection{Empirical Observations at Scale}


The formidable challenges of post-encoding analysis and rate control give rise to a critical question: \textit{is a reactive, frame-level controller intrinsically capable of guaranteeing perceptual quality?} Through large-scale measurements, we uncover three key empirical observations that serve as the foundation for viability in production cloud gaming scenarios.


\subsubsection{Observation 1: Feasibility of Reliable Per-Frame Quality Feedback}

The foundation of the QARC system lies in a reliable visual quality assessment method. Through an extensive user study covering four game types~\footnote{
    The selected game types include Multiplayer Online Battle Arena (MOBA), Sports Games (SPG), Role-Playing Games (RPG), and First-Person Shooters (FPS), representing a diverse spectrum of visual complexity, ranging from scenes with relatively smooth temporal evolution to highly dynamic competitive gameplay.
} (methodology detailed in Appendix~\ref{sec:user-test} and consistently applied throughout this work), in which we evaluated videos across a spectrum of quality levels against a 100~Mbps encoded lossless reference, we confirmed that traditional metrics such as VMAF and QP~\footnote{
    The Quantization Parameter (QP) regulates the compression granularity of macroblocks, serving as a crude proxy for visual quality. Constant QP (CQP) represents a quality-aware encoding strategy that is widely implemented in video encoders.
} are ill-suited for gaming content.

Our results reveal a significant misalignment between QP and subjective user ratings (detailed in Appendix~\ref{sec:app-qp}), suggesting that CQP is inadequate for direct application as a QARC solution in cloud gaming.
Moreover, as illustrated in Figure~\ref{fig:vmaf-and-sqoe}, VMAF fails to accurately capture perceptual degradation in gaming scenarios; for instance, frames achieving near-perfect VMAF scores of 99.5 can yield a normalized Differential Mean Opinion Score (DMOS) of only 95, falling short of perceptual losslessness. This discrepancy stems from VMAF's optimization for television-based video viewing, which limits its generalizability to PC monitors and gaming-specific graphics.
To bridge this gap, we leveraged large-scale real-world user feedback from the $T$ cloud gaming platform to construct a massive dataset, training a domain-specific model, SQoE, based on the DeepVQA~\cite{kim2018deep} architecture. Our subjective evaluations confirm that SQoE successfully captures the nuances of gaming visual quality, achieving high alignment with user ratings (Figure~\ref{fig:vmaf-and-sqoe}). Crucially, we quantified a definitive p-lossless threshold for gaming content at an SQoE score of $q^\star = 98.5$ (corresponding to 100 normalized DMOS), establishing that \textbf{\textit{maintaining an SQoE score above 98.5 effectively guarantees perceptual losslessness}}.


\subsubsection{Observation 2: Predominant Stability in Bitrate Demand with Transient 
Spikes} \label{sec:ob2}
\begin{figure}[t]
    \begin{minipage}[t]{0.48\linewidth}
        \includegraphics[width=\linewidth]{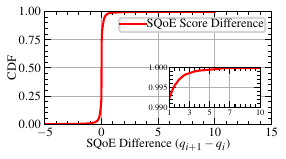}
        \caption{Distribution of SQoE score differences between consecutive frames.}
        \label{fig:stability}
    \end{minipage}
    \hfill
    \begin{minipage}[t]{0.48\linewidth}
        \includegraphics[width=\linewidth]{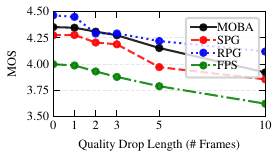}
        \caption{User ratings vs. quality drop duration for different game types.}
        \label{fig:big-drop-new}
    \end{minipage}
\end{figure}

As illustrated in Figure~\ref{fig:system-overview}, gaming scenes can be broadly categorized into phases of smooth, stable evolution interspersed with abrupt complexity spikes.
To quantitatively characterize the temporal dynamics of gameplay, we analyzed inter-frame SQoE variations across video clips captured from real-world game sessions.
As shown in Figure~\ref{fig:stability}, the distribution of these fluctuations is heavily concentrated near zero, with over 95\% falling within $\pm 0.2$. However, a pronounced tail reveals the presence of sporadic yet severe demand spikes, where differences can exceed 10.
This observation indicates that \textbf{\textit{reactive adjustment leveraging this stability suffices for the vast majority of the session}}, whereas \textbf{\textit{robust management of demand spikes is critical to preventing transient quality degradation}}. This dichotomy motivates us to decompose rate control into two distinct regimes.


\subsubsection{Observation 3: Tolerance to Single-Frame Glitches} \label{sec:big-drop}
To quantify the perceptual impact of transient artifacts, we established a low-quality threshold of SQoE $q_{\rm low} = 90$ (corresponding to < 90 normalized DMOS). We then systematically modulated the \emph{duration} of quality excursions below this threshold while maintaining surrounding content above the p-lossless target $q^\star$, collecting subjective 5-point Mean Opinion Scores (MOS) (Figure~\ref{fig:big-drop-new}).
The results reveal a sharp temporal threshold: degradation confined to a \textit{\textbf{single-frame interval}} exhibits negligible perceptual impact, whereas longer durations precipitate a consistent decline in user ratings across game types. This tolerance window effectively converts the inherent one-frame control lag into a manageable recovery budget, implying that \textbf{\textit{recovering from transient quality drops within a single-frame interval is sufficient to preserve perceptual losslessness}}.

\subsection{Our Insight} \label{sec:insight}
Building on the aforementioned challenges and empirical observations, we distill three pivotal design insights essential for deploying QARC in production cloud gaming systems. These insights address the unique constraints of the domain and guide the comprehensive design of SAIL, spanning \textit{perception}, \textit{control}, and \textit{transport} coordination, \textit{\textbf{to achieve bandwidth savings without compromising perceptual quality.}}

\subsubsection{Insight 1: Accurate VQA via Domain-Specialized Distillation} \label{sec:insight1}

Traditional QARC solutions typically rely on computationally intensive pixel-based models for visual quality analysis. In contrast, post-encoding analysis is restricted to sparse encoder information (e.g., QP, residual statistics) and must operate under strict latency budgets, necessitating the use of lightweight models. To ensure reliability despite these stringent constraints on input information and model capacity, we implement four complementary strategies to reinforce the VQA model in SAIL.

\noindent\textbf{(i) Knowledge Distillation.} We employ a knowledge distillation strategy, utilizing the sophisticated SQoE metric as a teacher model to equip the lightweight student model with robust baseline knowledge of gaming content.

\noindent\textbf{(ii) Game-Specific Fine-Tuning.} We train distinct models optimized for specific game types (e.g., MOBA, FPS). This specialization reduces the content variance the model must handle, thereby improving inference accuracy across diverse gaming scenarios.

Furthermore, SAIL’s estimator is decision-oriented rather than universally precise: it serves rate control by determining whether bitrate changes satisfy the p-lossless criterion. Accordingly, \textit{\textbf{accuracy is required primarily near $q^\star$}}, and \textit{\textbf{the cost of over-estimation dominates that of under-estimation}}. Leveraging this system-level insight, we introduce two additional safeguards:

\noindent\textbf{(iii) Asymmetric Training Objective.} We design an \textit{asymmetric loss function} that strictly penalizes over-estimation and prioritizes accuracy in the high-quality region.

\noindent\textbf{(iv) Error-Compensating Guardband.} We analyze the distribution of prediction errors and appropriately raise the quality control threshold. This compensatory adjustment effectively buffers against residual inaccuracies, ensuring system reliability.



\subsubsection{Insight 2: Hybrid Rate Control via Reactive Stabilization and Proactive Prevention}

Based on the observations detailed in \S~\ref{sec:ob2}, we decompose the rate control strategy into two distinct regimes: (i) reactive stabilization during steady states, and (ii) proactive prevention of transient degradation.

\noindent\textbf{(i) Reactive Stabilization.} In visually stable scenes, the temporal coherence of content renders historical feedback a reliable predictor of future complexity. Leveraging this stationarity, we utilize a history-based approach to model the bitrate-quality relationship and guide dynamic rate adjustments. However, recognizing the asymmetric risk profile--where bitrate under-allocation compromises user experience while over-allocation merely reduces bandwidth savings--we enforce an \textbf{\textit{asymmetric adjustment policy}}. This mechanism adopts a conservative, progressive approach for bitrate reduction to safely harvest bandwidth savings, while enabling aggressive, immediate updates for bitrate increments to rapidly rectify quality deficits.

\begin{figure}[t]
    \begin{minipage}[t]{0.48\linewidth}
        \includegraphics[width=\linewidth]{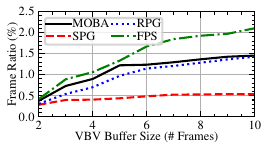}
        \caption{Relatively big frame ratio under different VBV buffer sizes.}
        \label{fig:background-vbv-size}
    \end{minipage}
    \hfill
    \begin{minipage}[t]{0.48\linewidth}
        \includegraphics[width=\linewidth]{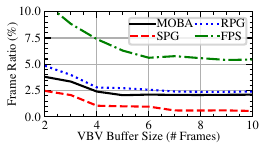}
        \caption{Low quality frame ratio under different VBV buffer sizes.}
        \label{fig:vbv-buffer-vmaf}
    \end{minipage}
\end{figure}

\noindent\textbf{(ii) Proactive Prevention.} We leverage the Video Buffer Verifier (VBV) to compensate for the inherent lag of reactive control. The VBV buffer acts as a critical mechanism within video encoders for temporal bitrate allocation. It functions as a virtual buffer that absorbs the constant target bitrate while allowing the encoder to draw bits flexibly to satisfy varying frame complexities. A larger buffer extends the window for resource allocation: on the one hand, it enables the encoder to borrow bits from simple frames to accommodate the high demands of complex scenes; on the other hand, it inevitably amplifies frame size fluctuations (Figures~\ref{fig:background-vbv-size} and~\ref{fig:vbv-buffer-vmaf}).
In conventional cloud gaming, where the encoding rate closely tracks the estimated network capacity, the VBV buffer is strictly limited to a small, static size (typically 3--5 frames) to prevent large frames from triggering network congestion~\cite{nvenc-vbv}.
SAIL redefines this paradigm by capitalizing on the \textit{bandwidth headroom} available when QARC reduces the encoding rate below network capacity. We \textit{\textbf{dynamically expand the VBV buffer}} based on this headroom, allowing the encoder to accommodate sudden bitrate surges during complexity spikes. This mechanism effectively absorbs bitrate demand fluctuations within the available bandwidth margin, thereby mitigating the transient quality degradation.

\subsubsection{Insight 3: Cross-Layer Coordination for CC and Accurate Bandwidth Estimation}
To ensure low latency and high reliability, we build our congestion control upon Pudica~\cite{Wang2024}, a state-of-the-art solution tailored for cloud gaming's system requirements.
Pudica maintains accurate control by estimating the BUR and dynamically switching between Multiplicative Increase (MI) and Additive Increase-Multiplicative Decrease (AI-MD) based on the current load. However, QARC introduces significant fluctuations in frame size and bitrate, which causes the actual encoding bitrate to fall far below Pudica's estimated bandwidth during stable periods. This sustained under-sending impairs CC by degrading its bandwidth probing capability, leading to the underestimation of available bandwidth. When scene complexity surges with higher bitrate demands, CC imposes restrictions on scene quality even if network capacity is sufficient.

To address this, SAIL designs a cross-layer coordination mechanism for \textit{\textbf{accurate bandwidth estimation}}. Specifically, it estimates not only the BUR under the current actual bitrate but also the network capacity, calculates the estimated BUR under the CC-dictated bitrate, and selects either the MI or AI-MD mechanism to determine the target CC bitrate. This mechanism prevents CC from underestimating available bandwidth, ensures network readiness for abrupt bitrate surges, and preserves low latency and efficient bandwidth utilization amid frame size fluctuations.

\section{Design of SAIL}

\begin{figure}[t!]
    \centering
    \includegraphics[width=\linewidth]{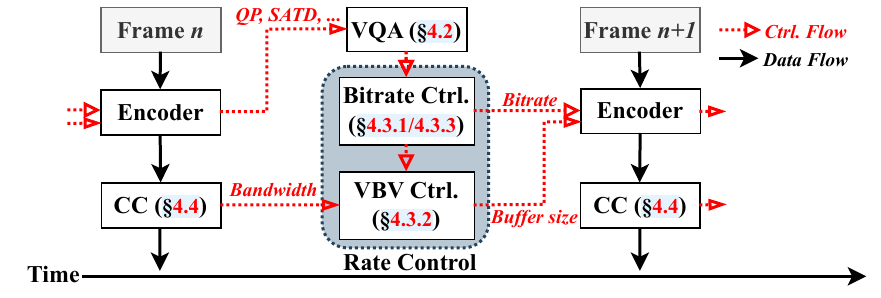}
    \caption{SAIL system overview.}
    \label{fig:cloud-gaming-qas}
\end{figure}

\subsection{System Overview}

Guided by the three design insights in \S~\ref{sec:insight}, SAIL operates as a \textit{post-encoding} analysis-and-control loop. It coordinates three key modules to maximize bandwidth efficiency while guaranteeing perceptual losslessness (Figure~\ref{fig:cloud-gaming-qas}):

\begin{enumerate}[wide, topsep=0pt, itemsep=0pt]
    \item \textbf{Encoder-Driven VQA (\S\ref{sec:design-vqa}):} Implementing \textit{Insight 1}, SAIL ensures reliable assessment under strict constraints. It synergizes knowledge distillation and asymmetric training to empower a lightweight student model, delivering accurate SQoE estimation using solely zero-cost encoder statistics (e.g., QP, residual statistics).
    
    \item \textbf{Hybrid Rate Control (\S\ref{sec:rate-control}):} Implementing \textit{Insight 2}, SAIL decomposes the complex control problem into \textit{reactive stabilization} and \textit{proactive prevention}. It employs an asymmetric policy to optimize steady-state bitrate allocation, while simultaneously orchestrating dynamic VBV adjustments to preemptively mitigate quality drops during complexity spikes.
    
    \item \textbf{Network-Aware Co-Design (\S\ref{sec:system-level-optimization}):} Following \textit{Insight 3}, SAIL prevents the transport layer from underestimating capacity during low-bitrate periods. By actively probing the network, it ensures the CC remains ready to accommodate sudden bitrate \textit{shoot-ups} for quality recovery.
\end{enumerate}

\subsection{Encoder-Driven VQA}
\label{sec:design-vqa}

\begin{figure}[t]
    \centering
    \begin{subfigure}[t]{0.48\linewidth}
        \centering
        \includegraphics[width=\linewidth]{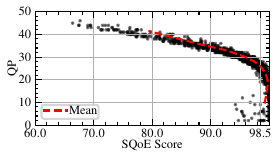}
    \end{subfigure}
    \begin{subfigure}[t]{0.48\linewidth}
        \centering
        \includegraphics[width=\linewidth]{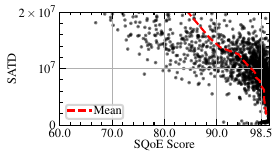}
    \end{subfigure}
    \caption{Relationship between QP/SATD and SQoE scores.}
    \label{fig:encoder-param-vs-vmaf}
\end{figure}

To achieve real-time quality assessment without heavy computational overhead, SAIL leverages the hardware encoder itself as a feature extractor.

\subsubsection{Zero-Cost Input Feature Selection}

Instead of processing pixel data, we extract signals directly from the encoder's internal state. These signals are \textit{zero-cost} because they are naturally computed during the compression process:

\begin{itemize}[wide, topsep=0pt, itemsep=0pt]
    \item \textbf{QP and Sum of Absolute Transformed Differences (SATD).} Reflect the quantization level and residual complexity, serving as primary indicators of spatial distortion.
    
    \item \textbf{No. of Motion Vector (MV).} The horizontal (MVx) and vertical (MVy) components of motion vectors, which represent the displacement of a block in the current frame relative to a reference frame. The number of MVx and MVy correlates with the smoothness of motion within the frame.

    \item \textbf{Intra Ratio.} The proportion of intra-frame prediction blocks (intra-blocks) to the total number of blocks. A high intra ratio indicates significant differences between the current and previous frames, often caused by rapid motion, scene changes, or cuts.

    \item \textbf{Encoded Frame Size.} The encoded frame size should also be included as an input, as motion vectors and intra ratio alone cannot independently reflect visual quality. For instance, a rapidly changing scene allocated a very high bitrate may not necessarily have worse quality than a smooth scene. Only under the constraint of similar bitrate allocation, i.e., comparable frame sizes, do higher motion vectors and intra ratio more strongly indicate potentially lower visual quality.
\end{itemize}

A detailed ablation study examining the impact of these input features is presented in Appendix~\ref{sec:app-vqa-input}.

\begin{figure}[t]
    \centering
    \begin{subfigure}[t]{0.48\linewidth}
        \centering
        \includegraphics[width=\linewidth]{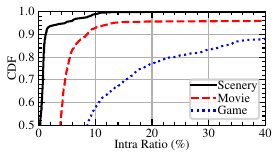}
    \end{subfigure}
    \begin{subfigure}[t]{0.48\linewidth}
        \centering
        \includegraphics[width=\linewidth]{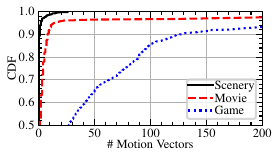}
    \end{subfigure}
    \caption{Intra ratio and motion vectors effectively capture content complexity.}
    \label{fig:natural-comparison}
\end{figure}

As illustrated in Figure~\ref{fig:encoder-param-vs-vmaf}, QP and SATD exhibit a strong correlation with SQoE, validating them as effective inputs. Furthermore, as shown in Figure~\ref{fig:natural-comparison}, MV and intra ratio effectively discriminate between simple and complex scenes, providing a robust basis for quality estimation.

\noindent\textbf{Input Preprocessing.} Given the significant disparity in the magnitude of input parameters--such as QP, which ranges from 0 to 51, and SATD, which is on the order of $10^7$--we normalized all inputs to ensure training stability. Furthermore, as shown in Figure~\ref{fig:natural-comparison}, the distributions of MVx, MVy, and intra ratio are highly long-tailed. To correct this skew, we applied a $\log(1 + x)$ transformation.

\subsubsection{Strategies for Reliability}

To ensure high reliability despite these stringent computational and latency constraints, we employ four synergistic strategies designed to reinforce model robustness.

\noindent\textbf{Knowledge Distillation.} To map these features to perceptual scores, we employ a knowledge distillation approach. A heavy, high-accuracy teacher model (SQoE) generates ground-truth quality scores offline. A lightweight student model--a simple fully-connected network--is then trained to predict these scores using only the encoder features.

\noindent\textbf{Game-Specific Fine-Tuning.} Since different games exhibit distinct feature distributions, a universal model often underperforms. We therefore adopt a \textit{pre-train-then-fine-tune} strategy: the student model is pre-trained on a diverse dataset and then fine-tuned on game-specific data to maximize accuracy for the target application.

A detailed ablation study examining the impact of these training strategies is presented in Appendix~\ref{sec:app-vqa-train}.

\noindent\textbf{Asymmetric Training Objective.} Guided by the insight in \S~\ref{sec:insight1}, we formulate the training loss as:

\begin{equation}
    \mathcal{L}(\boldsymbol{q}, \boldsymbol{\hat{q}}) = \dfrac{1}{N}\sum_{i = 1}^N \mu_i(q_i - \hat{q}_i)^2
\end{equation}

where $\boldsymbol{\hat{q}}$ and $\boldsymbol{q}$ denote the teacher's ground-truth and student's predicted SQoE scores, respectively. To enforce a conservative bias, we set the weight $\mu_i = 8$ when $q_i > \hat{q}_i > q_{\rm low}$ to penalize quality overestimation near the p-lossless threshold, and $\mu_i = 1$ otherwise.

An ablation study investigating the sensitivity of the parameter $\mu_i$ is presented in Appendix~\ref{sec:app-vqa-penalty}.

\noindent\textbf{Error-Compensating Guardband.} Upon completing model training and fine-tuning, extensive evaluations (detailed in \S~\ref{sec:vqa-exp}) reveal an average over-estimation error of approximately 0.2 across diverse game types. Consequently, we incorporate a safety margin by elevating the control threshold to $q^\star + 0.2$, thereby effectively compensating for potential prediction variance.

\begin{figure}[t]
    \begin{minipage}[t]{\linewidth}
        \includegraphics[width=\linewidth]{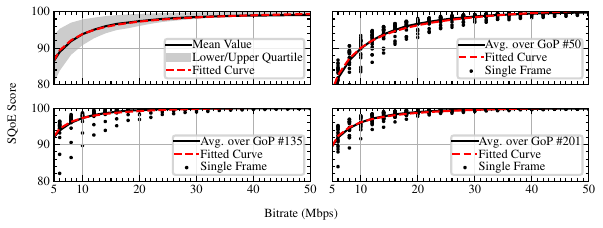}
        \caption{The inverse proportion relationship between bitrate and SQoE. Each subfigure corresponds to data sampled from a single Group of Pictures (GoP).}
        \label{fig:bitrate-vmaf}
    \end{minipage}
    \hfill
\end{figure}

\subsection{Hybrid Rate Control}
\label{sec:rate-control}

SAIL's rate control module jointly optimizes the encoding bitrate and VBV buffer size to navigate both steady-state evolution and abrupt scene transitions, maximizing bandwidth efficiency while preserving perceptual quality.


\subsubsection{Reactive Stabilization (Dynamic Bitrate Step)}

As illustrated in Figure~\ref{fig:bitrate-vmaf}, we formulate the bitrate-SQoE relationship as an inverse proportion function $q = q_{\rm max} - k_1 / (R + k_2)$.
Leveraging the mathematical properties of this formulation (detailed in Appendix~\ref{sec:derivation}), we analytically deduce that the online adjustment step is linearly related to the offline baseline, which can be derived via curve fitting on the collected video dataset.
This enables us to unify the complex scene dynamics into a single scaling parameter $\alpha_i$:
\begin{equation}
    \Delta R_i(q \to q_{\text{target}}) = \alpha_i \cdot \Delta R_{\text{offline}}(q \to q_{\text{target}})
\end{equation}

To balance rapid recovery with stable savings, we analytically modulate this predicted step $\Delta R_i$ with an asymmetric policy factor $M_i$, such that the final executed action is $\Delta R_{i, \text{online}} = M_i \cdot \Delta R_i$.

\noindent\textbf{Bitrate Increase.} Upon quality violation ($q_i < q^\star$), immediate restoration is imperative. We set the recovery target as $q_{\text{target}} = q^\star$ and apply a constant gain factor $M_i = A$:
\begin{equation}
    \Delta R_{i, \text{online}} = A \cdot \Delta R_i(q_i \to q_{\text{target}}), \quad q_i < q^\star
\end{equation}
where $A > 1$ (empirically 1.8) serves as a safety margin to accelerate convergence.

An ablation study regarding the determination of the safety margin $A$ is provided in Appendix~\ref{sec:app-rate-gain}.

\noindent\textbf{Bitrate Decrease.} Conversely, when quality suffices ($q_i > q^\star$), we reduce bitrate cautiously to prevent undershooting. We set a local probing target $q_{\text{target}} = q_i - \Delta q_{\text{step}}$ and the factor $M_i$ is proportional to the distance to $q^\star$:
\begin{equation}
    \Delta R_{i, \text{online}} = \underbrace{\frac{q_i - q^\star}{q_{\rm max} - q^\star}}_{M_i} \cdot \Delta R_i(q_i \to q_{\text{target}}), \quad q_i > q^\star
\end{equation}
This mechanism ensures the bitrate reduction rate diminishes as quality approaches $q^\star$, enhancing stability near the equilibrium.

\noindent\textbf{Smooth Online Adaptation.} Finally, the scaling factor $\alpha_i$ is recursively updated to track scene complexity. To account for the policy modulation $M_i$, the expected quality response is modeled as $M_i(q_{\text{target}} - q_i)$. We derive the instantaneous scaling $\alpha_{\text{new}}$ by normalizing the actual observed quality change $\Delta q_{\text{actual}}$ against this expectation, effectively decoupling intrinsic scene dynamics from control strategy. The periodic update is smoothed via Exponentially Weighted Moving Average (EWMA)~\cite{hunter1986exponentially} to ensure stability:
\begin{equation}
    \begin{aligned}
    \alpha_{\text{new}} &= \frac{\Delta q_{\text{actual}}}{M_i(q_{\text{target}} - q_i)} \cdot \alpha_i \\
    \alpha_{i + 1} &= \lambda\alpha_i + (1 - \lambda)\alpha_{\text{new}}
    \end{aligned}
    \label{eq:param-update}
\end{equation}
where the smoothing factor $\lambda$ is empirically set to $0.7$. To exclude anomalous feedback, adaptation is performed only when the observed quality change $\Delta q_{\text{actual}}$ has the same sign as the target change $q_{\text{target}} - q_i$.

An ablation study regarding the impact of the smoothing factor $\lambda$ is detailed in Appendix~\ref{sec:app-rate-ewma}.

\subsubsection{Proactive Prevention (Dynamic VBV)}

Beyond reactive rate adjustment, SAIL leverages the bitrate headroom to proactively relax VBV constraints, further stabilizing visual quality during complex scene dynamics.

\noindent\textbf{Dynamic Buffer Scaling.} By decoupling the encoding bitrate from the network capacity, SAIL creates headroom to expand the VBV buffer.
Since the internal VBV management of hardware encoders like NVENC is proprietary and not fully transparent, a theoretically optimal buffer size is difficult to derive. Instead, we employ a heuristic approach to determine the adaptive buffer size (in frames) as:

\begin{equation}
    N_{\rm VBV} = \left\lfloor N \cdot \frac CR \right\rfloor
\end{equation}

where $N$ is the baseline buffer size, $C$ denotes the network capacity, and $R$ is the target encoding bitrate.
This ensures that as the encoding bitrate falls below the network capacity, the VBV buffer expands to accommodate potential bitrate spikes. Conversely, under \textit{best-effort} conditions ($C = R$), the buffer size reverts to its default baseline.

\noindent\textbf{Underflow Mitigation.} While an enlarged buffer permits more flexible allocation, extreme scene complexity (e.g., scene cuts) may cause the encoder to deplete the buffer, risking underflow. SAIL monitors the target bitrate $R$ and actual frame size $s$ in real time. Given the frame rate $f$, the nominal frame size is $s_{\rm ideal} = R / f$. If the actual frame size exceeds $1.5 \times s_{\rm ideal}$, indicating a potential underflow, SAIL proactively resets the encoding bitrate to $R = s_{\rm actual} \cdot f$ to maintain buffer stability.

\subsubsection{Aggressive Recovery (Bitrate Shoot-Up)}

While dynamic VBV minimizes quality drops, it cannot eliminate them entirely. When a low-quality frame occurs (e.g., SQoE < $q_{\rm low}$), the system must restore visual quality immediately. Although this reactive adjustment occurs one frame after the detected drop, a single-frame quality glitch is generally imperceptible and tolerable, as discussed in \S~\ref{sec:big-drop}.

However, determining the exact bitrate required for such recovery is non-trivial; the bitrate needed to achieve a target SQoE varies significantly across scenes, making precise prediction difficult. To avoid under-allocation that would prolong quality degradation, SAIL employs a \textit{bitrate shoot-up} mechanism, which immediately raises the target bitrate to the maximum available capacity to ensure rapid restoration to perceptually lossless quality.


\subsection{Network-Aware Co-Design} \label{sec:system-level-optimization}

Bitrate control cannot operate in isolation from network conditions. A key challenge is that transmitting at low bitrates (when quality is high) can cause standard CC to underestimate available bandwidth, preventing the shoot-up when it is suddenly needed.

To address this, we adapted Pudica~\cite{Wang2024} to better align with our requirements. Pudica estimates the BUR under the current actual bitrate and selects either the MI or AI-MD mechanism based on the estimated BUR to determine the target CC bitrate. By obtaining the network's bandwidth capacity $C$, we can infer the BUR $B$ at the CC bitrate $R$ using the BUR $B_{a}$ measured at the actual bitrate $R_{a}$, i.e., $B = B_{a} + (R - R_{a})/C$.

To estimate the network capacity, we perform a bandwidth test during game start, setting this as the initial value for the estimated network capacity $C$. During the game, if the actual bitrate remains below $\alpha R$ for more than Pudica's BUR estimation window $T_{\rm wd}$, we initiate the capacity probing procedure. 
We transmit frames in an alternating sequence of burst and pace. The paced frames are used to calculate the BUR, while the burst frame's ACK rate is used as a capacity sample $C_{\rm sample}$ to update the estimated network capacity $C = \rho C_{\rm sample} + (1-\rho)C$. This approach allows SAIL to quickly increase the encoding bitrate to the available bandwidth even when transmitting at a lower actual bitrate.

\begin{figure*}[t]
  \centering
  \begin{subfigure}[t]{0.24\linewidth}
    \centering
    \includegraphics[width=\linewidth]{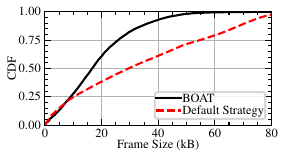}
    \caption{Average frame size}
    \label{fig:online-frame-size}
  \end{subfigure}
  \hfill
  \begin{subfigure}[t]{0.24\linewidth}
    \centering
    \includegraphics[width=\linewidth]{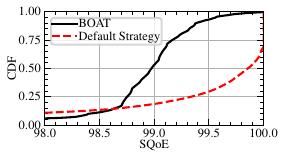}
    \caption{Average quality}
    \label{fig:online-quality}
  \end{subfigure}
  \hfill
  \begin{subfigure}[t]{0.24\linewidth}
    \centering
    \includegraphics[width=\linewidth]{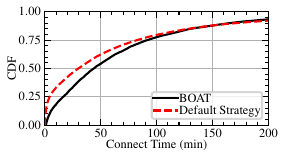}
    \caption{User engagement}
    \label{fig:online-subjective}
  \end{subfigure}
  \hfill
  \begin{subfigure}[t]{0.24\linewidth}
    \centering
    \includegraphics[width=\linewidth]{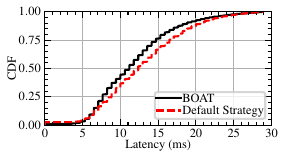}
    \caption{E2E latency}
    \label{fig:online-latency}
  \end{subfigure}
  
  \caption{Online deployment performance of SAIL comparing to $T$ cloud gaming platform default strategy.}
  \label{fig:online-overall}
\end{figure*}

\begin{figure*}[t]
  \centering
  \begin{subfigure}[t]{0.24\linewidth}
    \centering
    \includegraphics[width=\linewidth]{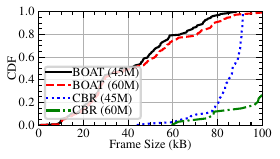}
    \caption{Average frame size}
    \label{fig:offline-frame-size}
  \end{subfigure}
  \hfill
  \begin{subfigure}[t]{0.24\linewidth}
    \centering
    \includegraphics[width=\linewidth]{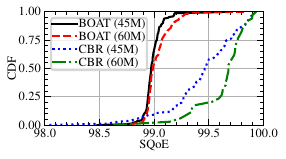}
    \caption{Average quality}
    \label{fig:offline-quality}
  \end{subfigure}
  \hfill
  \begin{subfigure}[t]{0.24\linewidth}
    \centering
    \includegraphics[width=\linewidth]{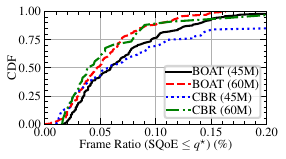}
    \caption{Low quality ratio}
    \label{fig:offline-low-quality}
  \end{subfigure}
  \hfill
  \begin{subfigure}[t]{0.24\linewidth}
    \centering
    \includegraphics[width=\linewidth]{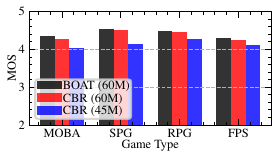}
    \caption{Subjective ratings}
    \label{fig:offline-subjective}
  \end{subfigure}
  \caption{Overall performance of SAIL comparing to CBR by offline simulation.}
  \label{fig:offline-overall}
\end{figure*}

\section{Experiments}

\subsection{Overall Performance}

\subsubsection{Online Deployment}

We deployed SAIL on the $T$ cloud gaming platform for a month-long large-scale evaluation, involving more than 57000 gaming sessions across 15 cities and 2 network types (Ethernet and Wi-Fi) with tens of thousands of subjective user ratings. Figure~\ref{fig:online-overall} illustrates the distributions of frame size, quality, user engagement, and E2E latency, with additional statistics summarized in Table~\ref{table:real-world}.

The results demonstrate that SAIL achieves a substantial \textbf{44.27\%} bandwidth savings (Figure~\ref{fig:online-frame-size}) by eliminating redundant bit allocation in stable scenes. Crucially, this efficiency gain is achieved without compromising visual fidelity; as shown in Figure~\ref{fig:online-quality}, the ratio of frames with quality exceeding the p-lossless threshold $q^\star = 98.5$ is higher than that of the default strategy. 

This improvement in quality-aware delivery translates directly into higher user satisfaction and retention. SAIL increases the average gaming session duration by \textbf{3.5~min}. As illustrated in Figure~\ref{fig:online-subjective}, users who are typically dissatisfied with the default strategy (exiting in under 3 minutes) show a marked tendency to stay longer when SAIL is active. Furthermore, the reduction in actual network data load yields systemic benefits: average E2E latency is decreased by \textbf{8.37\%} (Figure~\ref{fig:online-latency}), and the frequency of severe stall events is significantly mitigated (Table~\ref{table:real-world}), confirming that SAIL effectively resolves the bandwidth-latency trade-off inherent in best-effort systems.

\begin{center}
\begin{table}[]
\centering
\begin{tabular}{c|ccc}
\hline
     \textbf{Algo.}& \textbf{Avg. Q-MOS}& \textbf{Avg. L-MOS} & \begin{tabular}[c]{@{}c@{}}\textbf{90\%/95\%-tile}\\ \textbf{stall rate}\end{tabular} \\ \hline
SAIL & 3.05         &3.01                            & 0.19\%/1.21\%                                      \\ 
     Default  & 3.04   & 2.98                                              & 0.26\%/1.35\%                                                 \\\hline

\end{tabular}
\caption{Online evaluation of SAIL. \textit{Q-MOS} is short for \textit{Quality MOS} and \textit{L-MOS} is short for \textit{Latency MOS}.}
\label{table:real-world}
\end{table}
\end{center}

\vspace{-2.7em}

\subsubsection{Offline Evaluation}

To further evaluate the overall performance of SAIL, we conducted trace-driven simulations comparing it against the standard best-effort strategies used in current cloud gaming. Specifically, we implemented a CBR baseline with a fixed VBV buffer size to represent conventional production configurations, while SAIL enables dynamic buffer scaling. We evaluated SAIL and CBR under stable network conditions with constant capacities of 45~Mbps and 60~Mbps.

\noindent\textbf{Bandwidth Efficiency and Decoupling.} As illustrated in Figure~\ref{fig:offline-frame-size}, SAIL achieves significant bandwidth savings compared to CBR. While CBR consistently exhausts the allocated bitrate, SAIL reduces the average frame size by 40--50\%. Notably, we observe that SAIL's realized bitrate remains nearly constant even when the configured network capacity is adjusted. This demonstrates that SAIL successfully \textit{decouples} the encoding process from static caps, transitioning to a truly content-adaptive paradigm.

\noindent\textbf{Quality Stability and Consistency.} A key advantage of SAIL is its ability to deliver consistent visual quality across diverse content. As shown in Figure~\ref{fig:offline-quality}, while CBR's average quality fluctuates significantly between simple and complex video clips, SAIL maintains a nearly stable SQoE score ($\sim$ 99) regardless of the scene complexity. Furthermore, Figure~\ref{fig:offline-low-quality} shows that SAIL achieves a low-quality frame ratio comparable to high-bitrate CBR, yet does so with a much smaller bandwidth footprint.

\noindent\textbf{Perceptual Superiority.} The objective gains are further validated by our subjective user study. As reported in Figure~\ref{fig:offline-subjective}, users consistently assigned higher ratings to SAIL compared to CBR strategies with similar average bitrates. These results confirm that SAIL's reactive control and proactive prevention mechanisms collectively ensure a superior and more stable visual experience.

\subsection{VQA Model Evaluation} \label{sec:vqa-exp}

\subsubsection{Implementation}

We used a double-layer fully connected network and set the hidden layer dimension of the VQA model to 64. To train the model, we used the Adam optimizer~\cite{kingma2014adam} with a learning rate of 0.001. To prevent over-fitting, we divided the dataset into training, validation, and test sets in a 7:1:2 ratio and stopped training when the loss difference on the validation set fell below a certain threshold. The threshold was set to $10^{-5}$ during pre-training and $10^{-6}$ during fine-tuning. Both training and testing of the VQA model were conducted on an NVIDIA GeForce RTX 4070 GPU.




\subsubsection{Overall Performance}

Given that SAIL's VQA model is trained with an asymmetric objective tailored for rate control, its evaluation requires metrics beyond standard regression accuracy. In addition to the global RMSE and Pearson Linear Correlation Coefficient (PLCC)~\cite{pearson1896vii}, we report a biased RMSE (denoted as RMSE*) specifically within the high-SQoE interval $[q_{\rm low}, q_{\rm max}]$. This metric characterizes the model's reliability in the critical operational regime where precise, conservative estimation is essential for stable bitrate adaptation.


We adopted the following lightweight regression methods as baselines:
\begin{itemize}[wide, topsep=0pt, itemsep=0pt]
    \item \textbf{Polynomial Regression}: A degree-4 polynomial model.
    \item \textbf{Random Forest}~\cite{breiman2001random}: An ensemble learning method using 100 decision trees.
    \item \textbf{Support Vector Regression (SVR)}~\cite{smola2004tutorial}: Utilizing a Radial Basis Function (RBF) kernel.
    \item \textbf{XGBoost}~\cite{chen2016xgboost}: An optimized gradient boosting library with 100 estimators, trained using the same asymmetric loss function as SAIL for a fair comparison.
\end{itemize}

Table~\ref{table:vqa-compare} shows that SAIL significantly outperforms all baselines. SAIL achieves an RMSE of 0.7193, a 27.3--69.6\% reduction over Random Forest and XGBoost, and a PLCC of 0.9770. Notably, SAIL reduces the conservative estimation error (RMSE*) to 0.1486, marking a 70.9--77.8\% improvement. This superiority stems from SAIL's specialized student network, which better captures the non-linear mapping between encoder features and perceptual quality compared to general regression models. Furthermore, SAIL's training objective is tailored for asymmetric rate control tolerance, ensuring high precision in critical high-quality regimes where over-estimation would cause instability.



\vspace{0em}

\subsubsection{Computational Overhead}

After training the VQA model, we hard-coded its parameters in C++ language. We then ran the model $10^6$ times consecutively on an Intel(R) Core(TM) i7-9700 CPU @ 3.00GHz and measured the execution time. The average inference time per run was approximately $1.45 \times 10^{-2}$ ms, demonstrating that our VQA model fully meets the real-time requirements of cloud gaming.

\begin{figure*}[t]
  \centering
  \begin{minipage}[t]{0.74\textwidth}
    \centering
    \begin{minipage}[t]{0.33\linewidth}
      \centering
      \includegraphics[width=\linewidth,keepaspectratio]{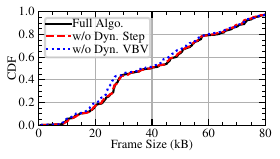}
    \end{minipage}%
    \hfill%
    \begin{minipage}[t]{0.33\linewidth}
      \centering
      \includegraphics[width=\linewidth,keepaspectratio]{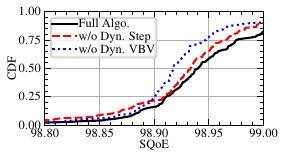}
    \end{minipage}
    \hfill%
    \begin{minipage}[t]{0.33\linewidth}
      \centering
      \includegraphics[width=\linewidth,keepaspectratio]{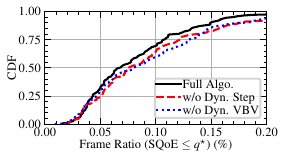}
    \end{minipage}
    \caption{Evaluation of components in the rate control algorithm.}
    \label{fig:module-ablation}
  \end{minipage}%
  \hspace{0.5em}%
  \begin{minipage}[t]{0.24\textwidth}
    \centering
    \begin{minipage}[t]{\linewidth}
      \centering
      \includegraphics[width=\linewidth,keepaspectratio]{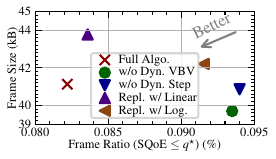}
    \end{minipage}%
    \caption{Evaluation of bitrate step fitting functions.}
    \label{fig:vsync-off}
  \end{minipage}
\end{figure*}

\begin{center}
\begin{table}[]
\centering
\begin{tabular}{c|cccc}
\hline
\textbf{Algorithm}                    & \textbf{RMSE}            & \textbf{PLCC}           & \textbf{RMSE*}           \\ \hline
Polynomial & 1.2724    & 0.9044      & 0.6702                    \\
Random Forest &0.9896&0.9439&0.5109\\
SVR &1.2120&0.9168&0.6304\\
XGBoost &2.3632&0.6631&0.6572\\
SAIL &\textbf{0.7193}&\textbf{0.9770}&\textbf{0.1486}\\ \hline
\end{tabular}
\caption{VQA model overall performance comparison.}
\label{table:vqa-compare}
\end{table}
\end{center}

\vspace{-2em}

\subsection{Micro-Benchmarks of Rate Control}

In this section, we investigate the impact of individual components of rate control algorithms on overall performance. As illustrated in Figure~\ref{fig:module-ablation}, we individually disabled two components, \textit{dynamic bitrate step} and \textit{dynamic VBV}, and compared the average encoded frame size and perceptual quality. These two components reduce the ratio of low-quality frames by 12.56\% and 12.08\% respectively while maintaining almost no additional bandwidth overhead (< 3\%).

To validate that the inverse proportionality model effectively captures the bitrate-SQoE relationship, we compared it against linear and logarithmic fitting for initial step calculation. As shown in Figure~\ref{fig:vsync-off}, our model achieves the lowest low-quality ratio (8.22\%) and the smallest average frame size (41.13~kB). In contrast, the linear model incurs a 6.45\% bandwidth overhead (43.79~kB) with a higher low-quality ratio (8.36\%), while the logarithmic model results in a 2.64\% larger frame size and a significantly worse low-quality ratio (9.15\%). These results demonstrate that the inverse proportionality model provides a more accurate mapping of quality sensitivity, allowing for aggressive yet precise bitrate reductions that avoid the over-allocation inherent in more conservative linear or logarithmic strategies.

\subsection{Network-Aware Co-Design}

In terms of CC, we utilized a specialized dummy client platform called Bonree~\cite{bonree} with millions of globally distributed end devices to simulate real user clients and evaluate our CC modifications.
The improved Pudica algorithm, referred to as Pudica*, was tested under extreme conditions by simulating a large bandwidth encoding-rate gap. The server transmitted video flow at a maximum of 10~Mbps bitrate for 1 second and then restore the maximum bitrate to 50~Mbps for 1 second. The bandwidth probed by CC during 10 Mbps streaming will determine the actual encoding rate when maximum bitrate is restored. Thus over-estimation will cause extra latency and even stall (E2E latency larger than 100~ms), while under-estimation will cause low bitrate. Such process is repeated for 5 minutes as one test session. This test reflects the bandwidth estimation performance of CCs when transmitting small volumes of data. As shown in Table~\ref{table:pudica-patch}, Pudica* achieves comparable bitrates, lower delay, and reduced stall rates.




\begin{center}
    \begin{table}[]
    \centering
    \begin{tabular}{c|cccc}
    \hline
    \textbf{Algo.} & \textbf{\makecell{Avg.\\ delay}} & \textbf{\makecell{95\%/99\% \\ -tile delay}} & \textbf{\makecell{Stall rate \\ $>$100/200~ms}} & \textbf{\makecell{Avg.\\ bitrate}}  \\ \hline

    
    Pudica & 17.4 & 25.7/42.5 & 0.32\%/0.58‰ & 45.6 \\
    Pudica* & 15.4 & 23.8/40.0 & 0.29\%/0.59‰ & 46.3 \\ \hline
    \end{tabular}
    \caption{Overall system-level performance of the CC at scale. Delays are measured in milliseconds (ms) and bitrate in megabits per second (Mbps).}
    \label{table:pudica-patch}
    \end{table}
\end{center}



\section{Related Work}


\subsection{Cloud Gaming System}

In a typical cloud gaming system, resource-intensive tasks such as game logic, graphics rendering, and other processes are offloaded to cloud servers. These servers receive user inputs, render game graphics in real-time, and stream them to user terminals via the network~\cite{DiDomenico2021,Alhilal2022,Lee2015}. 

According to prior research on the Quality of Experience (QoE) of cloud gaming, Motion-to-Photon (MTP) latency~\cite{9302797, jarschel2011evaluation} and video quality~\cite{slivar2016cloud, 8463416} are identified as primary factors significantly influencing user experience. To optimize video quality, Pudica~\cite{Wang2024} advanced congestion control algorithms, enabling more accurate link capacity prediction and improving final graphic quality. Regarding latency, methods such as multi-path transmission~\cite{Zhou2024} and optimizing time-consuming encoding and decoding queues~\cite{Meng2023} have been employed to minimize E2E latency in cloud gaming systems. 
However, prior work focusing on the visual quality of cloud gaming primarily addresses lower-layer characteristics, such as bitrate. These characteristics, however, do not fully align with human subjective perception of game visual quality in practice.

\subsection{Quality-Aware Rate Control}

\noindent\textbf{Video Quality Assessment.} Video quality assessment methods have evolved significantly alongside advancements in video processing technology. Initially, research focused on analyzing video frames using signal processing methods, employing features such as the SSIM~\cite{wang2004video} and its variants incorporating motion compensation~\cite{moorthy2010efficient} to describe quality. Subsequently, the development of feature extraction methodologies, such as those designed for analyzing local flicker distortion~\cite{choi2018video, choi2016flicker} or incorporating human visual perception simulations~\cite{zhang2015perception}, has made assessment algorithms more aligned with human perceptual logic. With the rise of deep learning, data-driven methods like DeepVQA~\cite{kim2018deep}, which uses CNN, and VMAF~\cite{vmaf}, which fuses multiple features, have gained widespread application. Recently, the strong video understanding capabilities of Large Multi-modal Models (LMM) have led to their application in this field, with relevant work including the construction of LMM evaluation benchmarks~\cite{liu2024tempcompass}. However, in the context of cloud gaming, the constraints of low latency and computational cost make it highly challenging to utilize raw images for video quality assessment. 

\noindent\textbf{QARC System.} QARC leverages video quality metrics to deliver stable, high-quality video content across various transmission scenarios. In VoD, where video content is available before user requests, preprocessing techniques such as pre-calculating video quality and integrating it into the QoE function~\cite{huang2019comyco,qin2019quality} are commonly employed. In RTC, the granularity of video quality assessment is relatively coarse, operating at the hundreds of milliseconds level~\cite{Huang2018, 10.1145/3474085.3475594, guo2017quality}. This allows for the use of complex feature extractors, such as heavy neural networks, to process image features~\cite{Huang2018, 10.1145/3474085.3475594}, or the inclusion of more historical frames in the assessment input~\cite{guo2017quality}. In contrast, cloud gaming, which involves real-time user interactions and exhibits more significant content changes than typical RTC scenarios, experiences non-negligible fluctuations in video quality at the frame level. This makes it difficult for RTC-based QARC methods to accurately assess quality.

\section{Conclusion}
In this paper, we presented SAIL, the first practical QARC solution tailored for the stringent constraints of cloud gaming. Starting from the critical challenges of latency and cost, we leveraged core observations regarding the viability of post-encoding control to break the efficiency bottleneck. SAIL materializes these insights into a holistic implementation that synergizes a zero-cost encoder-driven VQA model, a robust reactive rate control algorithm, and a network-aware congestion control mechanism. This design effectively decouples bitrate allocation from network capacity ensuring both quality and stability. Large-scale deployment demonstrates that SAIL achieves substantial bandwidth savings of 44.27\% and reduces E2E latency by 8.37\%, all while preserving p-lossless visual fidelity and enhancing user engagement. These results validate SAIL as a scalable solution that significantly improves the economic viability of cloud gaming platforms. Future work will explore extending this paradigm to broader real-time immersive applications.

\bibliographystyle{ACM-Reference-Format}
\bibliography{reference}

\appendix
\section*{Appendices}

\section{User Test Methodology} \label{sec:user-test}

\noindent\textbf{Demographics.} We recruited 250 participants (219 male, 31 female) from the $T$ cloud gaming platform. Participants were required to have at least one year of regular cloud gaming experience and self-reported normal or corrected-to-normal vision. Their ages ranged from 15 to 40 years (mean = 25.82, SD = 5.47). To encourage high-quality responses, participants who successfully completed the study were compensated with platform-specific virtual currency.

\noindent\textbf{Test Video Dataset.} The dataset for offline evaluation consisted of YUV-format video clips captured directly from the cloud gaming server during gameplay sessions conducted by an expert player. Each clip had a duration of 15 seconds, selected to encompass a broad spectrum of visual content ranging from stable scenes to high-velocity motion.

\noindent\textbf{User Test Process.} The study was conducted via a web-based survey interface accessible exclusively through desktop PC browsers. To ensure a varying baseline of display quality, we enforced a minimum screen resolution of 1080p and filtered out mobile user agents, though strict color calibration was not feasible in this remote setting. The stimuli were organized into batches using a Multi-Stimulus methodology. For each batch, participants could replay videos at will and were required to assign a 5-point MOS to each clip while also providing a preference ranking. The workflow proceeded sequentially to prevent bias.

\noindent\textbf{Invalid Feedback Filter.} To ensure data integrity, we implemented a rigorous filtering mechanism. First, we monitored the time spent on each video group; evaluations completed in less than 1.2$\times$ the cumulative video duration were flagged, and the participant was issued a warning. Second, we verified internal consistency by comparing MOS ratings with preference rankings using the Spearman Rank Order Correlation Coefficient (SROCC)~\cite{spearman1961proof}. Significant discrepancies also triggered a warning. Participants receiving three warnings were disqualified. In total, approximately 10 participants were excluded through this quality control process.

\begin{figure}[t]
    \begin{minipage}[t]{0.48\linewidth}
        \includegraphics[width=\linewidth]{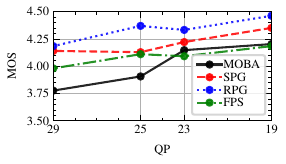}
        \caption{User ratings vs. QP levels.}
        \label{fig:app-qp}
    \end{minipage}
    \hfill
    \begin{minipage}[t]{0.48\linewidth}
        \includegraphics[width=\linewidth]{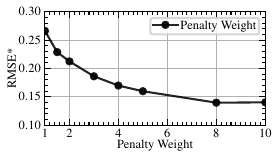}
        \caption{Ablation study of VQA penalty weight.}
        \label{fig:app-penalty}
    \end{minipage}
\end{figure}

\section{Relationship between QP and User Subjective Ratings} \label{sec:app-qp}

To evaluate the efficacy of the CQP strategy, we encoded test video clips across a range of QP levels and collected the corresponding subjective user ratings. As illustrated in Figure~\ref{fig:app-qp}, subjective ratings fail to consistently improve even as QP decreases. This indicates that QP alone is an unreliable proxy for actual perceptual quality. Consequently, CQP (and similar variants like CRF) are ill-suited for direct application as a QARC solution in cloud gaming. The fundamental limitation lies in the fact that QP is derived primarily from compression rate constraints, neglecting the perceptual impact of content complexity and motion dynamics. However, when integrated with other encoder-derived metrics and refined through specialized training strategies, QP can serve as a valuable feature enabling lightweight models to effectively mimic high-complexity VQA teachers.




\section{Derivation of Bitrate Control Step} \label{sec:derivation}

We model the relationship between encoding bitrate $R$ and SQoE score $q$ using an inverse function:
\begin{equation}
    q = q_{\rm max} - \dfrac{k_1}{R + k_2} \implies R = \dfrac{k_1}{q_{\rm max} - q} - k_2
    \label{eq:q-r-model}
\end{equation}

For a transition from current quality $q$ to target quality $q_{\rm target}$, the required bitrate adjustment $\Delta R$ is:
\begin{equation}
    \Delta R = R_{\rm target} - R = k_1 \left( \frac{1}{q_{\rm max} - q_{\rm target}} - \frac{1}{q_{\rm max} - q} \right)
\end{equation}

Applying the same functional form to the offline baseline dataset with parameters $k_{1, \text{offline}}$ and $k_{2, \text{offline}}$, the corresponding offline step $\Delta R_{\rm offline}$ is:
\begin{equation}
    \Delta R_{\rm offline} = k_{1, \text{offline}} \left( \frac{1}{q_{\rm max} - q_{\rm target}} - \frac{1}{q_{\rm max} - q} \right)
\end{equation}

Dividing the two yields:
\begin{equation}
    \dfrac{\Delta R}{\Delta R_{\rm offline}} = \dfrac{k_1}{k_{1, \text{offline}}}
\end{equation}

By defining the time-varying multiplier $\alpha_i := k_1 / k_{1, \text{offline}}$, we can express the online adjustment step as a simple scaling of the offline baseline: $\Delta R = \alpha_i \Delta R_{\rm offline}$.

\section{VQA Model Ablation Study}

\subsection{Input Parameters} \label{sec:app-vqa-input}

To assess the contribution of each input feature to the prediction accuracy, we normalized the parameter under investigation to a standard normal distribution in the dataset, followed by pre-training, fine-tuning, and testing. The ablation study results in Table~\ref{table:vqa-ablation} confirm that all selected features are indispensable, as removing any single parameter leads to a performance degradation of up to 25.6\% in RMSE*. Specifically, the removal of \textit{Intra Ratio} and \textit{MV} (Motion Vectors) causes the most significant accuracy drops in high-motion genres like SPG and FPS (increasing RMSE* by 17.8\% and 25.6\% respectively), as these features are critical for capturing temporal complexity and motion-induced artifacts. \textit{Frame Size} and \textit{QP} also prove vital across all categories, as they provide the direct signal for quantization-related distortion. These findings validate that our lightweight feature set effectively captures the multi-dimensional factors—ranging from spatial detail to temporal dynamics—that govern perceptual quality in interactive gaming streams.

\begin{center}
\begin{table}[t]
\centering
\begin{tabular}{c|cccc}
\hline
\textbf{\begin{tabular}[c]{@{}c@{}}Input\\ Parameter\end{tabular}} & \textbf{MOBA}   & \textbf{SPG}    & \textbf{RPG}    & \textbf{FPS}    \\ \hline
Original                                                           & \textbf{0.0953}&\textbf{0.1431}&\textbf{0.1510}&\textbf{0.2046} \\ \hline
QP                                                                 & 0.1034          & 0.1596          & 0.1636          & 0.2094          \\
Intra Ratio                                                        & 0.1011          & 0.1686          & 0.1607          & 0.2170          \\
SATD                                                               & 0.0981          & 0.1589          & 0.1614          & 0.2200          \\
MVx                                                                & 0.0969          & 0.1797          & 0.1630          & 0.2105          \\
MVy                                                                & 0.0966          & 0.1615          & 0.1599          & 0.2112          \\
Frame Size                                                         & 0.1055          & 0.1482          & 0.1564          & 0.2252          \\ \hline
\end{tabular}
\caption{VQA performance (RMSE*) after removing specific input parameters.}
\label{table:vqa-ablation}
\end{table}
\end{center}

\subsection{Training Strategies} \label{sec:app-vqa-train}

We employed three training strategies to validate the effectiveness of the pre-train-then-fine-tune approach:

\begin{itemize}[wide, topsep=0pt, itemsep=0pt]
\item\textbf{Pre-train:} the model is trained on the full dataset and directly evaluated on different game types.
\item\textbf{Fine-tune:} starting from the pre-trained model, it is further fine-tuned on game-specific datasets.
\item\textbf{Direct:} the model is trained from scratch solely on the dataset of the target game type.
\end{itemize}

The results in Table~\ref{table:vqa-acc} demonstrate that the fine-tuning strategy significantly enhances model accuracy across all game genres, achieving a 48.4\%--65.4\% reduction in RMSE* compared to the pre-training baseline. This substantial improvement indicates that while pre-training provides a broad foundational understanding of quality-feature mappings, fine-tuning effectively calibrates the model to the specific compression characteristics and visual complexities of different game types (e.g., FPS vs. MOBA). Furthermore, the fine-tuned model slightly outperforms the direct training approach (by 9.7\%--11.1\%), confirming that the knowledge transferred from the large-scale pre-training phase provides a more robust starting point than training from scratch, thereby improving generalization and final convergence.

\begin{center}
\begin{table}[]
\centering
\begin{tabular}{c|cccc}
\hline
\textbf{\begin{tabular}[c]{@{}c@{}}Training\\ Strategy\end{tabular}} & \textbf{MOBA} & \textbf{SPG} & \textbf{RPG} & \textbf{FPS} \\ \hline 
Pre-train& 0.1846&0.3850&0.4056&0.5921\\
Fine-tune&\textbf{0.0953}&\textbf{0.1431}&\textbf{0.1510}&\textbf{0.2046}\\
Direct&0.1055&0.1610&0.1702&0.2273\\ \hline
\end{tabular}
\caption{VQA performance (RMSE*) under different training strategies.}
\label{table:vqa-acc}
\end{table}
\end{center}

\subsection{Penalty Weight} \label{sec:app-vqa-penalty}

We analyzed the impact of the penalty weight $\mu$ on model performance, as illustrated in Figure~\ref{fig:app-penalty}. Increasing $\mu$ initially reduces the RMSE*; however, this improvement exhibits diminishing returns. Specifically, beyond $\mu = 8$, further increasing the penalty yields negligible performance gains. Consequently, we selected $\mu = 8$ as the optimal parameter for our final model configuration.

\begin{figure}[t]
    \begin{minipage}[t]{0.48\linewidth}
        \includegraphics[width=\linewidth]{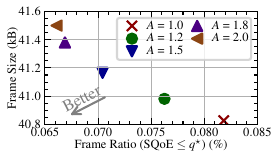}
        \caption{Ablation study of gain factor.}
        \label{fig:app-gain}
    \end{minipage}
    \hfill
    \begin{minipage}[t]{0.48\linewidth}
        \includegraphics[width=\linewidth]{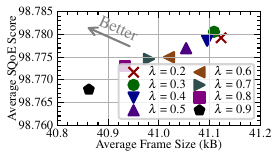}
        \caption{Ablation study of EWMA constant.}
        \label{fig:app-ewma}
    \end{minipage}
\end{figure}

\section{Rate Control Parameters Ablation Study}

\subsection{Gain Factor} \label{sec:app-rate-gain}

We encoded the test videos using varying gain factors to investigate the trade-off between quality stability and bandwidth consumption. As illustrated in Figure~\ref{fig:app-gain}, increasing the gain factor effectively suppresses the ratio of low-quality frames, albeit at the cost of increased average frame size. To prioritize quality assurance and minimize perceptible artifacts, we selected a more aggressive setting of $A = 1.8$ as the final operational parameter.

\subsection{EWMA Constant} \label{sec:app-rate-ewma}

As illustrated in Figure~\ref{fig:app-ewma}, the EWMA smoothing constant exerts a relatively minor influence on overall performance compared to other control parameters. Given that the observed variations in results were marginal across the tested range, we selected a balanced value of $\lambda = 0.7$ to ensure both stability and responsiveness.

\end{document}